\RequirePackage[l2tabu,orthodox]{nag}
\PassOptionsToPackage{hyphens,obeyspaces}{url}
\documentclass[manuscript,urlobeyspaces=true]{acmart}
\usepackage[utf8]{inputenc}

\usepackage{enumitem}
\usepackage{listings}
\usepackage{outlines}
\usepackage{subcaption}

\usepackage[repro]{reidpr}
\pdfid{}

\setlist[description]{leftmargin=\parindent,listparindent=\parindent}
\setlist[enumerate]{listparindent=\parindent}
\def\DefaultCutFileName{\def\CommentCutFile{\jobname.cut}}
\DefaultCutFileName
\makeatletter
\def\fps@figure{tbp}
\def\fps@table{tbp}
\makeatother

\newlength{\halffigwidth}
\copyrightyear{2024}
\acmYear{2024}
\setcopyright{rightsretained}

\begin{document}

\title{Zero-consistency root emulation for unprivileged container image build}

\author{Reid Priedhorsky}
\orcid{0000-0002-5348-0330}
\email{reidpr@lanl.gov}

\author{Michael Jennings}

\author{Megan Phinney}
\affiliation{%
  \department{High Performance Computing Division}
  \institution{Los Alamos National Laboratory}
  \city{Los Alamos}
  \state{NM}
  \country{USA}
}

\begin{abstract}

  Do Linux distribution package managers need the privileged operations they request to actually happen? Apparently not, at least for building container images for HPC applications. We use this observation to implement a root emulation mode using a Linux seccomp filter that intercepts some privileged system calls, does nothing, and returns success to the calling program. This approach provides no consistency whatsoever but appears sufficient to build all Dockerfiles we examined, simplifying fully-unprivileged workflows needed for HPC application containers.

\end{abstract}

\maketitle

\section{Introduction}

Scientific software for high performance computing (HPC) is increasingly deployed using Linux containers, which is a technology to package an application along with all its dependencies as a single unit called an \vocab{image}. A critical requirement for many HPC centers, including Los Alamos, is that user workflows must be fully unprivileged~\cite{priedhorsky2021privilege}; i.e., HPC users cannot be given elevated access of any kind to production resources. Within an unprivileged container, processes can have an effective user ID (EUID) of 0 (i.e., root) in a container and/or arbitrary capabilities, as well as access to some normally-privileged system calls, but this greater privilege is an illusion. Only unprivileged operations are actually available.

HPC container performance and reliability are typically best served by building images on the same supercomputer(s) targeted for deployment, due to their tightly specified architectures. This demands that building images, not just running them, must be fully unprivileged. However, build almost always uses traditional Linux distribution package managers such as \code{dpkg(8)}\footnote{Notation \code{foo(n)} indicates the thing named \code{foo} in man pages section \code{n}. §1 is user shell commands, and §2 is system calls, §8 is administrator commands~\cite{eaton2023man}.} or \code{rpm(8)} that assume they are running privileged, an assumption that has held for many years. While future package managers may relax this assumption, a build solution is needed for current distributions, which are in use now and may persist well beyond end-of-support when containerized.

One way to bridge this gap is \vocab{root emulation}, which replaces key privileged operations (e.g. \code{chown(2)} to change file ownership) with similar-enough unprivileged ones, thus fooling package managers into believing they are privileged. Existing approaches try to provide a consistent root-emulated environment. For example, record each \code{chown(2)} call and adjust the \code{struct stat} returned by later \code{stat(2)} calls so the process sees the same (fake) ownership that it set earlier. These tools must be either installed or bind-mounted into the container, and they add complexity and overhead while reducing compatibility.

Our insight is that usually \emph{consistency is not actually required} for building HPC application images. We can tell processes simple lies instead of complex ones. This paper describes a lightweight, non-consistent root emulation mode based on seccomp filters recently introduced in Charliecloud, LANL’s lightweight, fully unprivileged container implementation for HPC applications~\cite{priedhorsky2017sc}.\footnote{All authors are members of the Charliecloud team, and our arguments should be considered in that context.} We install a seccomp kernel filter that simply intercepts privileged system calls and returns success, without invoking the syscall or any user-space emulation of it. This remarkably unsophisticated root emulation appears workable for all images we tried; it is compatible with all distributions and libc’s as well as statically linked binaries; and it has no dependencies beyond a C compiler and the Linux kernel, not even libseccomp~\cite{moore2024libseccomp}.

\section{Fully unprivileged (Type~III) image build without root emulation?}

We previously proposed a tripartite classification of container implementations, based on the Linux namespaces used~\cite{kerrisk2013overview,kerrisk2013userns} and level of privilege needed to set up the container~\cite{priedhorsky2021privilege}:

\begin{itemize}

\item \inhead{Type~I} containers are the bare minimum, using the mount namespace but not the user namespace.\footnote{The other namespaces do not affect our classification.} They require privileged setup (root or \code{CAP_SYS_ADMIN}).

\item \inhead{Type~II} containers use mount and privileged user namespaces. They also require privileged setup (root or \code{CAP_SETUID} and \code{CAP_SETGID}). Many implementations call this type \vocab{rootless} because the main container processes are unprivileged, but we believe this is a misnomer because privileged helper programs (typically \url{newuidmap(1)} and \code{newgidmap(1)}) are needed to set up the container namespaces.

  \item \inhead{Type~III} containers use mount and unprivileged user namespaces; setup is unprivileged. \emph{Only Type~III containers are fully unprivileged throughout the container lifetime}. The benefit of Type~II over Type~III is greater flexibility of users and groups within the container.

\end{itemize}

\begin{figure}
  \begin{subcaptionblock}[T]{\halffigwidth}
\begin{lstlisting}
$ cat Dockerfile
FROM alpine:3.19
RUN apk add sl
$ ch-image build -t win --force=none .
  2. RUN.N apk add sl
updating existing image ...
fetch https://dl-cdn.alpinelinux.org/alpine/v3.19/m
fetch https://dl-cdn.alpinelinux.org/alpine/v3.19/c
(1/3) Installing ncurses-terminfo-base (6.4_p202311
(2/3) Installing libncursesw (6.4_p20231125-r0)
(3/3) Installing sl (5.02-r1)
Executing busybox-1.36.1-r15.trigger
OK: 8 MiB in 18 packages
grown in 2 instructions: win
\end{lstlisting}
    \phantomcaption
    \label{fig:naive-win}
  \end{subcaptionblock}%
  \hfill%
  \begin{subcaptionblock}[T]{\halffigwidth}
\begin{lstlisting}
$ cat Dockerfile.c7
FROM centos:7
RUN yum install -y openssh
$ ch-image build -t win --force=none .
  1* FROM centos:7
  2. RUN.N yum install -y openssh
[...]
  Installing : openssh-7.4p1-23.el7_9.x86_64    3/3
Error unpacking rpm package openssh-7.4p1-23.el7_9.
error: unpacking [...] failed [...]: cpio: chown
[...]
something went wrong, rolling back ...
[...]
error: build failed: RUN command exited with 1
\end{lstlisting}
    \phantomcaption
    \label{fig:naive-fail}
  \end{subcaptionblock}%
  \caption[]{Example Dockerfiles built with a Type~III (fully unprivileged) implementation and no root emulation. \subref{fig:naive-win}~succeeded because no privileged system calls were used, while \subref{fig:naive-fail}~failed because \protect\code{rpm(8)} tried to change a file’s owner, a privileged operation.}
  \label{fig:naive}
\end{figure}

It would be convenient for HPC if images could be built in a Type~III container naïvely, with no root emulation or other special measures. This does sometimes work. Figure~\ref{fig:naive-win} shows an example Dockerfile that builds with no root emulation; \code{apk(8)} can install \code{sl(1)} [sic] with no privileged system calls. On the other hand, in Figure~\ref{fig:naive-fail}, \code{rpm(8)} tried to change a file’s ownership with \code{chown(2)}, a privileged operation disallowed in an unprivileged container despite being container root.

This is why we need root emulation. What if \code{chown(2)} was not really \code{chown(2)} but rather an unprivileged substitute?

\section{Related work}

Charliecloud is not the first to implement root emulation, whether complex or simple. However, to our knowledge, in the context of container image builders, it was the first to provide complex emulation in 2020~\cite{priedhorsky2021privilege}, and it is now the first to provide simple emulation in 2023. This section details existing root emulation work, with a focus on image build.

\subsection{\protect\code{fakeroot(1)}}

\code{fakeroot(1)} is a program to run a command in a root-emulated environment. It is not a perfect simulation but rather just enough to work for its intended purpose, which is building distribution packages, allowing “users to create archives (tar, ar, .deb etc.) with files in them with root permissions/ownership”~\cite{man2021fakeroot}. There are at least three \code{fakeroot(1)} implementations~\cite[Table~1]{priedhorsky2021privilege} that hook processes in two different ways. \code{LD_PRELOAD} is a userspace mechanism that lets \code{fakeroot(1)} intercept shared library function calls (not syscalls); this is architecture-independent but cannot wrap statically linked executables. \code{ptrace(2)} is a kernel mechanism that can intercept system calls (and do many other things). It is architecture-dependent but can wrap statically linked executables. All \code{fakeroot(s)}s maintain state in order to provide a consistent emulated environment (e.g., so \code{stat(2)} is consistent with prior \code{chown(2)}), with a daemon and/or disk files.

Charliecloud was the first to implement \code{fakeroot(1)} injection into container image builds, as detailed in ~\cite{priedhorsky2021privilege}. It does this by installing \code{fakeroot(1)} into the image from package repositories of the containerized distribution, which requires detailed configuration for each supported distribution.


Singularity~\cite{kurtzer2017singularity}, like Charliecloud, is a container implementation targeting HPC applications. It also supports image build, though via “definition files” rather than the de facto standard Dockerfile~\cite{apptainer2023userguide}. The project forked in 2021~\cite{apptainer2021community,sylabs2022singularityce}. One of the forks, Apptainer, also supports root emulation via \code{fakeroot(1)}, based on Charliecloud with but with a key difference: the host’s \code{fakeroot(1)} is bind-mounted into the container~\cite{dykstra2022apptainer}. This trades the need to install it in the image for tighter dependence between the host and container libc, but it does not address the other drawbacks of \code{fakeroot(1)}.

\subsection{PRoot}

Another stand-alone root emulator is PRoot, which uses \code{ptrace(2)} to intercept system calls~\cite{vincent2022proot}, avoiding libc compatibility issues and allowing the tool to wrap static executables~\cite{trudgian2022proot}. PRoot can in fact use seccomp filters, but for a different purpose than Charliecloud: it is a performance optimization that lets PRoot avoid being notified about system calls it doesn’t need to intercept. However, the fundamental constraint of a complex, state-maintaining tool remain.

The other fork of Singularity, SingularityCE~\cite{sylabs2022singularityce}, bind-mounts the host’s \code{proot(1)} to provide root emulation for image build, building on the lessons of \code{fakeroot(1)} by Charliecloud and Apptainer~\cite{trudgian2022proot} for an arguably better implementation of complex root emulation.

\subsection{\protect\code{fakechroot(1)}}

Like some implementations of \code{fakeroot(1)}, \code{fakechroot(1)} uses \code{LD_PRELOAD} to intercept libc function calls, the main goal being to provide an unprivileged \code{chroot(2)}~\cite{roszatycki2019fakechroot}. It also provides a simple root emulation by substituting \code{/bin/true} for a configurable set of executables. This is sufficient to e.g.\ bootstrap a Debian distribution, but this emulation surface of executables only isn’t broad enough for general image building.

\section{Seccomp (filter mode)}

In this paper we are concerned with the “filter mode” seccomp,\footnote{Seccomp is short for \vocab{secure computing}, though its scope has expanded considerably since naming.} introduced in Linux 3.5 in 2012~\cite{kerrisk2024seccomp}. This lets a process install a filter to manipulate the system calls of itself and its children. This filter is a Berkeley Packet Filter (BPF)\footnote{As the name implies, BPF was originally designed to manipulate network packets but likewise has been expanded in scope.} program run by a kernel, and once installed it cannot be removed, i.e., it binds program children whether they like it or not. Notably, BPF does not have loops, so it can be verified for completion by the kernel.

The BPF filter is run by the kernel upon each system call. It has four inputs: (1)~the system call number (not name!), which varies by architecture; (2)~the syscall’s arguments, (3)~the current architecture (which can vary even within a process), and (4)~the instruction pointer. An important limitation is that BPF filters cannot dereference pointers. After its computation, the filter returns the disposition of the system call, which falls into three classes:
\begin{enumerate}

\item \inhead{Do not execute the syscall}, and one of (a)~kill the thread (Linux 3.5), (b)~kill the process (4.14), (c)~send SIGSYS to the thread (3.5), or (d)~return a specified \code{errno} (3.5).

\item \inhead{Execute the syscall}, and (e)~log it first (4.14) or simply (f)~execute it normally (3.5).

\item \inhead{Delete the decision to a userspace process}, either with (g)~\code{ptrace(2)} (3.5) or (h)~a file descriptor (5.0), which then chooses disposition a–f.

\end{enumerate}





Several container implementations use seccomp filters. Docker and Podman/Buildah have a filter specification feature, apparently intended as a simple allow/denylist for syscalls~\cite{docker2023seccomp}. distrobuilder (part of the LXC/LXD package) has a filter configuration language that could likely create root emulation like Charliecloud’s, but because users “must be root in order to run the distrobuilder tool”~\cite{distrobuilder-docs}, that potential capability is moot. Firejail is a tool to sandbox processes using container-like technologies such as namespaces; it does filter system calls using seccomp but not for root emulation~\cite{firejail2015features}. NsJail is a “light-weight process isolation tool” that uses namespaces~\cite{swiecki2024nsjail}.\footnote{Though it resides in Google’s GitHub organization, NsJail’s readme states that it “is NOT an official Google product” [emphasis in original].} It provides a seccomp configuration language that appears flexible enough to implement root emulation, but we are unaware of anyone having done so. Finally and notably, Enroot is a small container runtime that \emph{does} provide a lightweight root-emulation seccomp filter with the same approach as Charliecloud’s: “[w]e use a seccomp filter to trap all setuid-related syscalls, to make them succeed”~\cite{abecassis2020distributed}. However, the filter is less complete than Charliecloud’s, and Enroot does not provide a build capability, which is where the main root emulation challenge lies.

\section{Charliecloud’s seccomp filter}


Charliecloud zero-consistency root emulation installs a seccomp filter to intercept certain privileged system calls and fake their success. In pseudocode, our filter is:
\begin{quotation}
\begin{lstlisting}
if (privileged system call):
   do nothing
   return success
\end{lstlisting}
\end{quotation}
That is, from the point of view of a filtered process, these system calls always succeed, but if the process does anything to verify the actions requested, it will see that nothing happened.


\begin{figure}
  \centering
  \begin{minipage}{\halffigwidth}
\begin{lstlisting}
$ cat Dockerfile
FROM centos:7
RUN yum install -y openssh
$ ch-image build -t win -f Dockerfile.c7 .
  1* FROM centos:7
  2. RUN.S yum install -y openssh
[...]
  Installing : openssh-7.4p1-23.el7_9.x86_64    3/3
[...]
Complete!
--force=seccomp: modified 0 RUN instructions
grown in 2 instructions: win
\end{lstlisting}
  \end{minipage}
  \caption{Successful seccomp root-emulation build of the Dockerfile in Figure~\ref{fig:naive-fail} above.}
  \label{fig:seccomp-win}
\end{figure}

The 29 privileged syscalls we filter fall into four classes:
\begin{enumerate}

\item \inhead{File ownership} (7 syscalls): \code{chown(2)}, \code{fchownat(2)}, etc.

\item \inhead{User/group/capability manipulation} (19): \code{setresuid(2)}, \code{capset(2)}, etc.

\item \inhead{\code{mknod(2)} and \code{mknodat(2)}} (2): These two syscalls can be privileged or not. We must examine the file type argument before faking success (device file) or allowing the syscall (other types).

\item \inhead{Self-test} (1): \code{kexec_load(2)} reboots into a new kernel and is unlikely to ever be needed by HPC applications, so we use it to validate the filter after installation.

\end{enumerate}

Charliecloud’s source code has a table listing the numbers for each syscall on each of the six supported architectures,\footnote{Some syscalls are not implemented on all architectures; for example, arm64 lacks \code{chown(2)}, relying on user-space code to translate its calls to \code{fchownat(2)} instead.} and we translate this into a BPF program and install it with two C functions totalling about 150 lines of code including comments. Figure~\ref{fig:seccomp-win} shows a successful Charliecloud build using the zero-consistency root emulation mode of the Dockerfile introduced earlier in Figure~\ref{fig:naive-fail}.

An exception to the assumption that package managers don’t care about consistency is Debian’s \code{apt(8)}, which by default drops privileges for downloading packages over HTTP(S) and also \emph{verifies that they were dropped correctly}. This validation fails under our seccomp filter. We work around the problem awkwardly by detecting \code{apt(8)} and \code{apt-get(8)} in \code{RUN} instructions and injecting \code{-o APT::Sandbox::User=root} into their command lines, which disables privilege dropping for download.

\section{Discussion}

This paper presents a novel root emulation mode based on the principle that distribution package managers and similar tools do not need their privileged requests to be actually carried out, but rather are satisfied to be simply told what they want to hear (for the purpose of container image build at least). That is, we have implemented a seccomp filter for some privileged system calls that, instead of executing the syscall, does nothing and returns success to the userspace program. While limited to identity and files, this simple, zero-consistency root emulation is sufficient to build almost all the container images we tried. (Known exceptions are builds that call \code{unminimize(8)} or trigger certain systemd scripts, and these both seem to be implementation hassles rather than something fundamental about our approach.)

Alternately, one can use a complex, consistent root emulation using \code{fakeroot(1)} or \code{proot(1)}. Simple and complex both allow image build with a fully unprivileged Type~III container implementation, a critical requirement for HPC application containers. In our assessment, however, the simple, inconsistent seccomp method of root emulation has a number of advantages:

\begin{enumerate}

\item \inhead{Overhead.} The seccomp method imposes a relatively light overhead~\cite{larabel2020seccomp,zatoichi2017zatoichi} of its filter on every system call (not just those filtered), while the consistent method requires user-space emulation of system calls, making an extra program and possibly its shared libraries available to the container build, and state maintenance using a daemon process.

\item \inhead{Simplicity.} The seccomp method has no user-space component and does nothing to actually emulate any system calls; further, “emulation” is complete once the filter is installed (though see \code{apt(8)} workaround above). Also, because it does not maintain state, the seccomp method also intercepts fewer system calls.

\item \inhead{Compatibility.} The seccomp method is agnostic to libc and static/dynamic linking, and mostly agnostic to distribution, the exception being \code{apt(8)} above, though these properties are shared by PRoot and \code{fakeroot(1)} implementations based on \code{ptrace(2)}. Fewer intercepted system calls and no syscalls actually emulated has compatibility benefit as well.

\end{enumerate}

Importantly, however, the complex emulation is consistent on dimensions relevant to package management; a process under emulation can make changes to identity or privileged file metadata and have the emulated changes reflected back later. This seems to usually not matter for image build, but sometimes it does, e.g., \code{apt(1)} above.

Future work includes (1)~an optional wider set of emulated syscalls, such as \code{setxattr(2)}, which may allow systemd to be installed;\footnote{You might ask: “Why do I want systemd in my containers?” Indeed, you probably don’t, but it tends to pulled in as a dependency.} (2)~evaluate adding \emph{just a little} consistency, for user and groups IDs only, to remove the workaround for \code{apt(8)} explained above; and (3)~performance testing.


\bibliographystyle{ACM-Reference-Format}
\bibliography{refs}


\begin{thebibliography}{23}


\ifx \showCODEN    \undefined \def \showCODEN     #1{\unskip}     \fi
\ifx \showDOI      \undefined \def \showDOI       #1{#1}\fi
\ifx \showISBNx    \undefined \def \showISBNx     #1{\unskip}     \fi
\ifx \showISBNxiii \undefined \def \showISBNxiii  #1{\unskip}     \fi
\ifx \showISSN     \undefined \def \showISSN      #1{\unskip}     \fi
\ifx \showLCCN     \undefined \def \showLCCN      #1{\unskip}     \fi
\ifx \shownote     \undefined \def \shownote      #1{#1}          \fi
\ifx \showarticletitle \undefined \def \showarticletitle #1{#1}   \fi
\ifx \showURL      \undefined \def \showURL       {\relax}        \fi
\providecommand\bibfield[2]{#2}
\providecommand\bibinfo[2]{#2}
\providecommand\natexlab[1]{#1}
\providecommand\showeprint[2][]{arXiv:#2}

\bibitem[fir(2015)]%
        {firejail2015features}
 \bibinfo{year}{2015}\natexlab{}.
\newblock \bibinfo{title}{Features}.
\newblock
\newblock
\urldef\tempurl%
\url{https://firejail.wordpress.com/features-3}
\showURL{%
\tempurl}


\bibitem[Abecassis and Calmels(2020)]%
        {abecassis2020distributed}
\bibfield{author}{\bibinfo{person}{Felix Abecassis} {and}
  \bibinfo{person}{Jonathan Calmels}.} \bibinfo{year}{2020}\natexlab{}.
\newblock \bibinfo{title}{Distributed {{HPC}} applications with unprivileged
  containers}.
\newblock
\newblock
\urldef\tempurl%
\url{https://archive.fosdem.org/2020/schedule/event/containers_hpc_unprivileged/}
\showURL{%
\tempurl}


\bibitem[{Apptainer project}(2021)]%
        {apptainer2021community}
\bibfield{author}{\bibinfo{person}{{Apptainer project}}.}
  \bibinfo{year}{2021}\natexlab{}.
\newblock \bibinfo{title}{Community announcement}.
\newblock
\newblock
\urldef\tempurl%
\url{https://apptainer.org/news/community-announcement-20211130/}
\showURL{%
\tempurl}


\bibitem[{Apptainer project}(2023)]%
        {apptainer2023userguide}
\bibfield{author}{\bibinfo{person}{{Apptainer project}}.}
  \bibinfo{year}{2023}\natexlab{}.
\newblock \bibinfo{title}{Apptainer user guide}.
\newblock
\newblock
\urldef\tempurl%
\url{https://apptainer.org/docs/user/main/security.html#}
\showURL{%
\tempurl}


\bibitem[Dassen et~al\mbox{.}(2021)]%
        {man2021fakeroot}
\bibfield{author}{\bibinfo{person}{J.H.M. Dassen}, \bibinfo{person}{joost
  {witteveen}}, {and} \bibinfo{person}{Clint Adams}.}
  \bibinfo{year}{2021}\natexlab{}.
\newblock \bibinfo{booktitle}{\emph{fakeroot(1)}}.
\newblock \bibinfo{type}{Man page}.
\newblock
\urldef\tempurl%
\url{https://manpages.debian.org/bullseye/fakeroot/fakeroot.1.en.html}
\showURL{%
\tempurl}


\bibitem[{distrobuilder contributors}(2023)]%
        {distrobuilder-docs}
\bibfield{author}{\bibinfo{person}{{distrobuilder contributors}}.}
  \bibinfo{year}{2023}\natexlab{}.
\newblock \bibinfo{title}{distrobuilder documentation}.
\newblock
\newblock
\urldef\tempurl%
\url{https://linuxcontainers.org/distrobuilder/docs/latest/}
\showURL{%
\tempurl}


\bibitem[{Docker Inc.}(2023)]%
        {docker2023seccomp}
\bibfield{author}{\bibinfo{person}{{Docker Inc.}}}
  \bibinfo{year}{2023}\natexlab{}.
\newblock \bibinfo{title}{Seccomp security profiles for {{Docker}}}.
\newblock
\newblock
\urldef\tempurl%
\url{https://docs.docker.com/engine/security/seccomp/}
\showURL{%
\tempurl}


\bibitem[Dykstra(2022)]%
        {dykstra2022apptainer}
\bibfield{author}{\bibinfo{person}{Dave Dykstra}.}
  \bibinfo{year}{2022}\natexlab{}.
\newblock \bibinfo{title}{Apptainer without {{Setuid}}}.
\newblock
\newblock
\urldef\tempurl%
\url{https://doi.org/10.48550/arXiv.2208.12106}
\showDOI{\tempurl}
\showeprint[arxiv]{2208.12106}~[cs]


\bibitem[Eaton et~al\mbox{.}(2023)]%
        {eaton2023man}
\bibfield{author}{\bibinfo{person}{John~W. Eaton}, \bibinfo{person}{Rik Faith},
  \bibinfo{person}{G. WIlford}, \bibinfo{person}{Fabrizio Polacco}, {and}
  \bibinfo{person}{Colin Waton}.} \bibinfo{year}{2023}\natexlab{}.
\newblock \bibinfo{booktitle}{\emph{man(1)}}.
\newblock \bibinfo{type}{Man page}.
\newblock
\urldef\tempurl%
\url{https://man7.org/linux/man-pages/man1/man.1.html}
\showURL{%
\tempurl}


\bibitem[Kerrisk(2013a)]%
        {kerrisk2013overview}
\bibfield{author}{\bibinfo{person}{Michael Kerrisk}.}
  \bibinfo{year}{2013}\natexlab{a}.
\newblock \showarticletitle{Namespaces in operation, part 1: {{Namespaces}}
  overview}.
\newblock \bibinfo{journal}{\emph{Linux Weekly News}} (\bibinfo{date}{Jan.}
  \bibinfo{year}{2013}).
\newblock
\urldef\tempurl%
\url{https://lwn.net/Articles/531114/}
\showURL{%
\tempurl}


\bibitem[Kerrisk(2013b)]%
        {kerrisk2013userns}
\bibfield{author}{\bibinfo{person}{Michael Kerrisk}.}
  \bibinfo{year}{2013}\natexlab{b}.
\newblock \showarticletitle{Namespaces in operation, part 5: {{User}}
  namespaces}.
\newblock \bibinfo{journal}{\emph{Linux Weekly News}} (\bibinfo{date}{Feb.}
  \bibinfo{year}{2013}).
\newblock
\urldef\tempurl%
\url{https://lwn.net/Articles/532593/}
\showURL{%
\tempurl}


\bibitem[Kerrisk(2024)]%
        {kerrisk2024seccomp}
\bibfield{author}{\bibinfo{person}{Michael Kerrisk}.}
  \bibinfo{year}{2024}\natexlab{}.
\newblock \bibinfo{title}{Seccomp}.
\newblock
\newblock
\urldef\tempurl%
\url{https://man7.org/training/download/splc_seccomp_slides-mkerrisk-man7.org.pdf}
\showURL{%
\tempurl}


\bibitem[Kurtzer et~al\mbox{.}(2017)]%
        {kurtzer2017singularity}
\bibfield{author}{\bibinfo{person}{Gregory~M. Kurtzer},
  \bibinfo{person}{Vanessa Sochat}, {and} \bibinfo{person}{Michael~W. Bauer}.}
  \bibinfo{year}{2017}\natexlab{}.
\newblock \showarticletitle{Singularity: {{Scientific}} containers for mobility
  of compute}.
\newblock \bibinfo{journal}{\emph{PLOS ONE}} \bibinfo{volume}{12},
  \bibinfo{number}{5} (\bibinfo{date}{May} \bibinfo{year}{2017}).
\newblock
\showISSN{1932-6203}
\urldef\tempurl%
\url{https://doi.org/10.1371/journal.pone.0177459}
\showDOI{\tempurl}


\bibitem[Larabel(2020)]%
        {larabel2020seccomp}
\bibfield{author}{\bibinfo{person}{Michael Larabel}.}
  \bibinfo{year}{2020}\natexlab{}.
\newblock \bibinfo{title}{Seccomp filters get a very nice speed-up with
  {{Linux}} 5.11}.
\newblock
\newblock
\urldef\tempurl%
\url{https://www.phoronix.com/news/Linux-5.11-SECCOMP-Performance}
\showURL{%
\tempurl}


\bibitem[Moore and {others}(2024)]%
        {moore2024libseccomp}
\bibfield{author}{\bibinfo{person}{Paul Moore} {and}
  \bibinfo{person}{{others}}.} \bibinfo{year}{2024}\natexlab{}.
\newblock \bibinfo{title}{libseccomp}.
\newblock \bibinfo{howpublished}{The libseccomp Project}.
\newblock
\urldef\tempurl%
\url{https://github.com/seccomp/libseccomp}
\showURL{%
\tempurl}


\bibitem[Priedhorsky et~al\mbox{.}(2021)]%
        {priedhorsky2021privilege}
\bibfield{author}{\bibinfo{person}{Reid Priedhorsky}, \bibinfo{person}{R.~Shane
  Canon}, \bibinfo{person}{Timothy Randles}, {and} \bibinfo{person}{Andrew~J.
  Younge}.} \bibinfo{year}{2021}\natexlab{}.
\newblock \showarticletitle{Minimizing privilege for building {{HPC}}
  containers}. In \bibinfo{booktitle}{\emph{Proc. {{SC}}}}.
\newblock
\urldef\tempurl%
\url{https://doi.org/10.1145/3458817.3476187}
\showDOI{\tempurl}


\bibitem[Priedhorsky and Randles(2017)]%
        {priedhorsky2017sc}
\bibfield{author}{\bibinfo{person}{Reid Priedhorsky} {and} \bibinfo{person}{Tim
  Randles}.} \bibinfo{year}{2017}\natexlab{}.
\newblock \showarticletitle{Charliecloud: {{Unprivileged}} containers for
  user-defined software stacks in {{HPC}}}. In
  \bibinfo{booktitle}{\emph{Supercomputing}}.
\newblock
\urldef\tempurl%
\url{https://doi.org/10.1145/3126908.3126925}
\showDOI{\tempurl}


\bibitem[Roszatycki(2019)]%
        {roszatycki2019fakechroot}
\bibfield{author}{\bibinfo{person}{Piotr Roszatycki}.}
  \bibinfo{year}{2019}\natexlab{}.
\newblock \bibinfo{title}{fakechroot}.
\newblock
\newblock
\urldef\tempurl%
\url{https://github.com/dex4er/fakechroot/blob/2.20.1/man/fakechroot.pod}
\showURL{%
\tempurl}


\bibitem[Swiecki et~al\mbox{.}(2024)]%
        {swiecki2024nsjail}
\bibfield{author}{\bibinfo{person}{Robert Swiecki} {et~al\mbox{.}}}
  \bibinfo{year}{2024}\natexlab{}.
\newblock \bibinfo{title}{nsjail}.
\newblock
\newblock
\urldef\tempurl%
\url{https://github.com/google/nsjail}
\showURL{%
\tempurl}


\bibitem[{Sylabs Inc.}(2022)]%
        {sylabs2022singularityce}
\bibfield{author}{\bibinfo{person}{{Sylabs Inc.}}}
  \bibinfo{year}{2022}\natexlab{}.
\newblock \bibinfo{title}{{{SingularityCE}} is {{Singularity}}}.
\newblock
\newblock
\urldef\tempurl%
\url{https://sylabs.io/2022/06/singularityce-is-singularity/}
\showURL{%
\tempurl}


\bibitem[Trudgian(2022)]%
        {trudgian2022proot}
\bibfield{author}{\bibinfo{person}{Dave Trudgian}.}
  \bibinfo{year}{2022}\natexlab{}.
\newblock \bibinfo{title}{proot based non-root / non --fakeroot builds}.
\newblock
\newblock
\urldef\tempurl%
\url{https://github.com/sylabs/singularity/issues/880}
\showURL{%
\tempurl}


\bibitem[Vincent et~al\mbox{.}(2022)]%
        {vincent2022proot}
\bibfield{author}{\bibinfo{person}{Cédric Vincent} {et~al\mbox{.}}}
  \bibinfo{year}{2022}\natexlab{}.
\newblock \bibinfo{title}{{{PRoot}} — chroot, mount --bind, and binfmt\_misc
  without privilege/setup}.
\newblock
\newblock
\urldef\tempurl%
\url{https://proot-me.github.io/}
\showURL{%
\tempurl}


\bibitem[{Zatoichi}(2017)]%
        {zatoichi2017zatoichi}
\bibfield{author}{\bibinfo{person}{{Zatoichi}}.}
  \bibinfo{year}{2017}\natexlab{}.
\newblock \bibinfo{title}{Zatoichi's {{Engineering Blog}}}.
\newblock
\newblock
\urldef\tempurl%
\url{https://zatoichi-engineer.github.io/2017/11/06/seccomp-bpf.html}
\showURL{%
\tempurl}


\end{thebibliography}

\end{document}